\documentclass{elsarticle}

\usepackage{geometry}                
\usepackage{graphicx}
\usepackage{amssymb}
\usepackage{epstopdf}
\usepackage{natbib}
\bibpunct{(}{)}{;}{a}{}{,}

\begin{document}

\title{Target Space $\neq$ Space\\\ \\Nick Huggett\\Department of Philosophy M/C 267\\University of Illinois at Chicago\\Chicago, IL 60613\\\ \\huggett@uic.edu\\+1-312-996-3022 (phone)\\+1-312-413-2093 (fax)}

\maketitle

\newpage

\noindent\textbf{Abstract:} This paper investigates the significance of T-duality in string theory: the indistinguishability with respect to all observables, of models attributing radically different radii to space -- larger than the observable universe, or far smaller than the Planck length, say. Two interpretational branch points are identified and discussed. First, whether duals are physically equivalent or not: by considering a duality of the familiar simple harmonic oscillator, I argue that they are. Unlike the oscillator, there are no measurements `outside' string theory that could distinguish the duals. Second, whether duals agree or disagree on the radius of `target space', the space in which strings evolve according to string theory. I argue for the latter position, because the alternative leaves it unknown what the radius is. Since duals are physically equivalent yet disagree on the radius of target space, it follows that the radius is indeterminate between them. Using an analysis of \cite{BraVaf:89}, I explain why -- even so -- space is observed to have a determinate, large radius. The conclusion is that observed, `phenomenal' space is not target space, since a space cannot have both a determinate and indeterminate radius: instead phenomenal space must be a higher-level phenomenon, not fundamental.\\


\section{T-Duality}

Consider a closed, classical bosonic string in a Minkowski spacetime with a compact spatial dimension, $x$, of radius $R$.\footnote{My technical presentation follows, especially, \citet[237]{Gre:99}, \citet[Ch 17]{Zwi:04} and \cite{BraVaf:89}.} 

\begin{figure}[htbp] 
   \centering
   \includegraphics[width=2in]{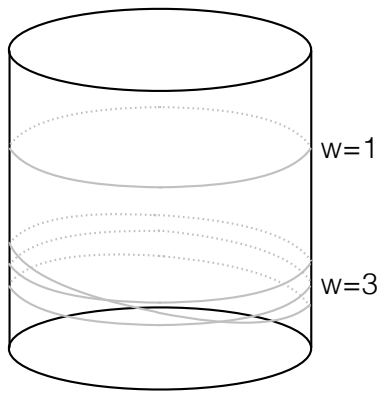} 
   \caption{Two closed strings in a 2-dimensional space with one compact dimension. One string is wrapped once around $x$, and the other three times -- winding numbers $w=1$ and $w=3$ respectively.}
   \label{fig:wound}
\end{figure}

Adopting the conventions that the string coordinate $0\leq\sigma\leq\pi$, while the compact spatial coordinate $0\leq x\leq2\pi$, the state of the string with respect to $x$ is 

\begin{equation}
\label{eq:wound}
X(\sigma,\tau) = 2w\sigma R + \ell_s^2p\tau + \mathrm{vibrational\ terms}.
\end{equation}
$X(\sigma,\tau)$ is the $x$-coordinate of the point of the string worldsheet with worldsheet coordinates $(\sigma,\tau)$: hence the state of a string specifies an embedding of the worldsheet into spacetime. The first term describes the $w$-fold winding of the string: for instance, if the string is wound once around $x$, so $w=1$, then $X$ ranges from 0 to $2\pi$ as $\sigma$ ranges from 0 to $\pi$. The second term represents the linear momentum; the constant $\ell_s$ is the `characteristic string length'. For simplicity, we shall ignore vibrations, since they do not change the substance of this paper. 

In wisely chosen string coordinates, substituting $X$ in the Hamiltonian gives

\begin{equation}
\label{eq:HamP}
H =  \frac{T}{2\pi}\int d\sigma \ \dot{X}^2 + X'^2 =  \frac{T}{2\pi}\int d\sigma\ (\ell_s^2p)^2 + (wR)^2,
\end{equation}
where $T$ is another constant, the string tension. Not surprisingly there is a kinetic term, plus a term from the winding, or stretching, of the string. The next step is to quantize.

Momentum first. The closed dimension implies a periodic boundary condition for momentum eigenstates $\Psi_k(x)=e^{ikx}$ (ignoring normalization, and setting $\hbar=1$)

\begin{equation}
\label{ }
\Psi_k(0) = \Psi_k(2\pi R) \Rightarrow e^{ik\cdot 0} = 1 = e^{ik\cdot2\pi R} \Rightarrow k=0, \pm1/R, \pm 2/R\dots.
\end{equation}
In other words, momentum is quantized: $|k|=n/R$, with `wave number' $n$. Substituting into the Hamiltonian (\ref{eq:HamP}), we obtain the spectrum

\begin{eqnarray}
\label{eq:Hspec}
E_{n,w} & = & \frac{T}{2\pi}\int d\sigma\ (\ell_s^2n/R)^2 + (wR)^2.
\end{eqnarray}

Now winding. Assuming interactions, in QM a string can change the number of times it is wound around a closed dimension. Hence $w$ is not a constant, classical c-number of the system, but  a dynamical \emph{quantum} quantity, described by a wavefunction. (Pay close attention to this point, as it is crucial:  because the winding number can change over time, a quantum string can be in \emph{superposition} of states of different winding numbers. Without this move there is no T-duality -- it is a quantum phenomenon.) The winding term in (\ref{eq:HamP}) depends on $l=wR$, which must have a discrete spectrum since $w$ does. Thus much as before, these eigenstates have the form $\Phi_l(y)=e^{ily}$ around a circle with coordinate $y$, but with radius $1/R$. In that case the periodic boundary condition $\Phi_l(0)=\Phi_l(2\pi/R)$ yields

\begin{equation}
\label{ }
e^{il\cdot0} = 1 = e^{il\cdot2\pi/R} \Rightarrow  l= 0, \pm R, \pm2R, \dots = wR,
\end{equation}
as required. In summary, the state of a quantum string involves (the tensor product of) two wavefunctions, one representing its position/momentum, and another representing its winding.

The question is of course, `where is the circle on which the $\Phi_l(y)$ wavefunction lives?' It can't just be in space, because then $\Phi_l(y)$  describes a second object which we could expect to find somewhere. Instead, there must be a new `internal' dimension associated with each  compact dimension of space; hence Witten calls $y$  `another ``direction'' peculiar to string theory' \cite[29]{wit96}. His proposal is not that \emph{space} has an extra dimension for every dimension a string can wrap around, but rather that treating winding as a quantum observable means that it can be represented like momentum on a \emph{non-spatial} circle. Or more precisely, when we consider the space of all states of any momentum or winding, we find two quantum `position' operators, $x$ and $y$, respectively corresponding to position in physical space (radius $R$) and in a new `winding space' (radius $1/R$). But observables represent physical quantities, so we have to take both `positions' and spaces seriously, even if only one is physical space -- let's call the other `winding space'. But remember, the string winds around physical space, while the winding number wave lives in winding space.

\ 

\setlength{\leftskip}{1cm}

\noindent Semi-technical aside: as usual, $x$ and $y$ are `position' operators for physical and winding space. Moreover, as $\hat p=-i\partial/\partial x$ is the momentum observable with eigenvalues $k=n/R$ in the periodic plane wave states $e^{inx/R}$, so $\hat w=-i\partial/\partial y$ is the winding observable with eigenvalues $l=wR$ in the winding states $e^{iwRy}$. Thus each space is associated with identical canonical commutation relations, $[\hat x,\hat p]=i$ and $[\hat y,\hat w]=i$ (the observables from different spaces commute). Therefore, since position and momentum generate the algebra of observables, each space has, formally speaking, exactly the same observables, individuated as functions of $x$ and $\hat p$ or $y$ and $\hat w$.\footnote{Of course there are observables involving operators from both the spaces, but since the latter commute, such observables are always the commutative product of a pair of observables, one from each space. So all the points we need go through trivially, and we will ignore them.} This correspondence will be important below.

\setlength{\leftskip}{0cm}

\ 

Such internal spaces are familiar -- spin states and gauge field states, for instance -- so there is nothing new yet. But look again at (\ref{eq:Hspec}), the spectrum of the Hamiltonian. It's easy to check that a string with wave number $n$ and winding number $w$ in a space with radius $R$ has the same eigenvalue as a string with \emph{winding} number $n$ and \emph{wave} number $w$, \emph{but which lives in a space of radius} $\ell_s^2/R$: 

\begin{equation}
\label{eqn:T-D}
n \leftrightarrow w \quad \mathrm{and} \quad R \to  \ell_s^2/R
\end{equation}
The  second string has a spatial wavefunction in a compact dimension of radius $\ell_s^2/R$, and hence -- by the same reasoning as before -- a winding wavefunction that in a compact dimension of reciprocal radius, namely $R/\ell_s^2$.\footnote{This quantity appears to have units of $length^{-1}$ but it should be understood as involving multiplication by a unit area to give it the correct units. A similar point applies elsewhere that quantities appear to have the wrong units.} If the first string lives in a space with radius $R>\ell_s$, then the second string lives in a space of radius $\ell_s^2/R<\ell_s$: the strings are `reflected' through $\ell_s$.

Then ($i$) because they have the same Hamiltonian, both strings will have the same mass spectrum (because in string theory the Hamiltonian controls the mass). Moreover, ($ii$) because the roles of momentum and winding are reversed in the Hamiltonian by (\ref{eqn:T-D}), the dynamics of the spatial wavefunction in one string become the dynamics of the winding wavefunction in the other, and \textit{vice versa} -- in other words, the duality between momentum and winding is preserved over time.

Now, because wavefunctions in physical space are in exact correspondence with the wavefunctions in winding space, every observable pertaining to physical space corresponds to an observable pertaining to winding space -- related to winding just as the former is related to momentum. (And \textit{vice versa}.) Then, since momentum and winding are exchanged by (\ref{eqn:T-D}), every observable pertaining to physical space is exchanged with a corresponding observable pertaining to winding space -- and because the wavefunctions are also exchanged, the value of the new observable will equal that of the original observable for the first string. (And \textit{vice versa}.) In short, the pattern of observable quantities will be preserved by (\ref{eqn:T-D}) -- all that changes is whether the quantity is understood to pertain to physical or to winding space.\footnote{The mapping introduces a $\ell_s^2$ factor, but these  can be absorbed in a trivial rescaling of observables, so we will ignore it.}

\ 
\setlength{\leftskip}{1cm}

Continuation of the technical aside: a little more formally, the point is that the algebra of observables on spatial wavefunctions for one string is mapped onto the identical (as we saw above) algebra of observables on winding wavefunctions of the other -- with $x\leftrightarrow y$ and $\hat p\leftrightarrow\hat w$. Since the wavefunctions are also exchanged, the values associated with \emph{all} corresponding elements of the algebra of observables are preserved by (\ref{eqn:T-D}) -- the entire pattern of expectation values.

\setlength{\leftskip}{0cm}

\ 

Thus the systems are equivalent: the Hamiltonian and hence dynamics are the same, and all physical quantities agree (making the standard assumption that the observables exhaust the physical quantities of a theory). This equivalence, and others comparably strong, are known as `dualities', and the two theories related by it are `dual' to each other, or `duals'. In particular, (\ref{eqn:T-D}) is known as `T-duality', where -- depending on whom you ask -- `T' either stands for `target space' (i.e., the space in which the string lives), or for `torus'. For a further discussion of the relation of T-duality to other dualities, see Polchinski (this volume).

The significance of T-duality will be the subject of the next section, but for now a concrete example, taken from \citet{BraVaf:89}, will help illustrate some of its implications. Take T-duals, $T_1$ and $T_2$, which differ in the radius that they postulate for a closed spatial dimension: a circumference of $10^{12}$ light years (two orders of magnitude bigger than the visible universe) on the one hand, and $10^{-94}$m on the other (so assuming a value for the characteristic string length of $10^{-33}$m, two orders of magnitude above the Planck length). Thus $T_1$ and $T_2$ (apparently) make radically different assertions about the size of a spatial dimension. Before T-duality, one would assume that simple observations would rather readily choose between them, but that can't be right if the duals are physically equivalent.

To understand how the equivalence manifests itself, Brandenberger and Vafa consider an archetypical measurement of the radius: fire off a particle of known velocity -- a photon, say -- and time its journey around space. Suppose the result is a trillion years: that seems pretty conclusive evidence for the large radius story, $T_1$. But, in terms of $T_1$, how does the measurement work? The photon has a spatial wavefunction $\psi(x,t)$, which evolves, according to the Hamiltonian, from being localized nearby, via a journey away of $10^{12}$ years, to being localized nearby again. However, $T_2$ can also account for this result. 

The photon is a low energy mode of the string -- the easiest thing we can excite. In $T_1$, such a state involves momentum excitations: $k=n/R$, so momentum is inverse to the radius of space, and a very large radius means very low momentum, hence low energy. But in $T_2$, even the smallest, $n=1$, excitation has huge momentum, while the energy of stretching a string around a dimension of radius $10^{-94}$m is tiny. Thus in $T_2$ the photon in the experiment is described by a winding state. And of course the states are indeed dual: a $T_1$ momentum state maps to a $T_2$ winding state, the former with wave number, and the latter with winding number, $n$. Thus, according to $T_2$ the photon has the dual state represented by the same wavefunction $\psi(y,t)$ \emph{but in winding space}. Then, because the Hamiltonian is the same, but with the roles of physical and winding space reversed, the photon evolves in exactly the same way -- namely a `journey' around \emph{winding} space, taking $10^{12}$ years, as observed. No surprise there: in $T_2$ physical space is tiny, hence winding space, with reciprocal radius is huge. Note that this analysis shows that because the experiment is characterized as timing a low energy particle (of a given type), by stipulation it involves a process in the larger of physical and winding space, and so is guaranteed to take a long time, \emph{guaranteed} to produce the phenomena of a large radius space in either dual. (Indeed, because they are dual, the process is guaranteed to take the same time, be an observation of the same large radius.)

And the equivalence generalizes. Any process in \emph{physical and winding} space according to $T_1$, corresponds to a process in $T_2$ in \emph{winding and physical} space, and so no measurements or observations of even the most hypothetical kind will distinguish them. And so, the question comes up of which we should take to be correct -- or indeed, whether the difference between tiny and huge is a true physical one at all. In the next section we will review the main responses to this situation, and see reasons to favor one.

\section{T-Duality and the Nature of Space in String Theory} 

\subsection{A Taxonomy of Interpretations}

T-duality provides a mapping between a pair of theories that agree (under the mapping) on the expectation values of all observables in all states, and on the evolutions of all states.\footnote{It's important to note a mismatch between a philosopher's and a physicist's use of `theory' at this point. Roughly, physicists distinguish theories by the mathematical form of the canonical variables, while philosophers often further distinguish them by the values of their constants. In physicists' but not philosophers' terms, T-duals are the same theory, and T-duality is a `self-duality', but there is no substantive disagreement about the facts. I have decided to follow philosophers' usage here, because it will facilitate the following discussion to emphasize the differences between the duals.\\\indent Also note that \cite{Matsubara:2013uq} and \cite{Read:2014fk} provide alternative complementary categorizations of the stances on duality, useful in the context of different  background philosophical issues. \citet{DieDon:15} proposes a distinction similar to my second taxonomic fork.} It's important to note that `observables' here does not have any narrow philosophical empiricist meaning: it denotes the collection of hermitian operators (subject to any selection rules), \emph{not} some `special' collection of properties to which we have especially `direct' access. Indeed, the observables are thus those operators normally thought of as representing the totality of physical, quantum mechanical quantities. And with respect to those quantities the theories are -- under the mapping -- in perfect agreement.

The core of the mapping was given above, partly in (\ref{eqn:T-D}): 

\begin{eqnarray}
\label{eq:map}
\nonumber n & \leftrightarrow & w\\
 R & \to & \ell_s^2/R \\
 \nonumber \Psi(x,t) \to \Phi(x,t) & \mathrm{and} & \Phi(y,t) \to \Psi(y,t)
\end{eqnarray}
\textit{Prima facie}, in one system a string has momentum $n/R$, and is wound $w$ times around a dimension of radius $R$. In the other, it has momentum $Rw/\ell_s^2$ and is wound $n$ times around a dimension of radius $R'\equiv\ell_s^2/R$. And in the quantum mechanical treatment spatial and winding-spatial parts of the wavefunction are interchanged (which, in the case of simultaneous momentum and winding \emph{eigen}states, entails $n \leftrightarrow w$).

Normally one thinks of c-numbers such as $R$ as also physical, in which case the duals describe different physical situations. But normally, c-number parameters can be determined by the values of quantum quantities: the charge on the electron, say, by scattering probabilities. A duality means they cannot be so determined by the values of the observables of the theory: the pattern of expectation values is preserved. So we should at least leave open that such differences in the c-numbers do not, after all, represent physical differences. Below, invoking a simple duality, I will argue that indeed they do not. But for now we have our first interpretive decision: either the T-duals agree on the physical world or they do not. If they do, then for the purposes of this enquiry they say the same, \emph{all in}; we will not be interested in alleged non-physical differences. 

Commentators have been pretty uniform in taking the stance that the T-duals should indeed be taken as giving the same physical description: especially, see \cite{Dawid:2007gf}, \cite{Matsubara:2013uq}, \cite{Rickles:2011db}, and \cite{Rickles:2013ul}.  I argue below that they are correct. Plausibly it is also the implicit view of string theorists in general (for instance, \citealt[247]{Gre:99}), though their words sometimes seem ambiguous on the point: especially, \cite{Teh:2013rc} identifies remarks suggesting that one dual may be more fundamental than another. However, taking the view that duals are physically equivalent, a second interpretive decision awaits.\footnote{I want to thank Dave Baker for making me see that there are distinct options here.}

To describe the options now facing us, it is first necessary to be more careful in distinguishing the different conceptions of space that have entered the discussion. First, there is, what I shall call, `\emph{phenomenal space}'. `Phenomenal' here does not denote some narrow philosophical restriction to what is immediately given by the senses, but rather contrasts `theoretical'. That is, when a new, more fundamental scientific theory explains an established, less fundamental theory, which has stood the test of experiment, then we can speak of the latter as the `phenomena' \emph{relative} to the former. Being relative, the distinction is suited for the historical process by which today's novel `theory' becomes experimentally vindicated, and eventually becomes tomorrow's bedrock empirical given: Kepler's laws were phenomena for Newton's, but Newton's laws were phenomena for general relativity.\footnote{It is not essential for the reader to accept this historical picture. It's helpful to accept a theory-phenomenon distinction, which has its origins in \citet[1-19]{Car:83} and \citet{BogWoo:88}. What I take from them is the idea that `phenomena' are abstracted from direct observation events, and so have a `theoretical' structure themselves: I then add that a theory might therefore be phenomena relative to a more fundamental theory. Even this point is not essential to the paper, as I state explicitly to what concepts `phenomenal space' applies: but the term is loaded, and so it was necessary to explicate its philosophical context.} In the present case, the more fundamental theory is string theory, which aims to explain, amongst other things, our current account of space. This account is expressed in part in our current best scientific theories, quantum field theory and general relativity, in the small (high energy physics) and large (cosmology): these are the `phenomena' relative to string theory, and we will thus call the space that they describe `phenomenal'. But `phenomenal space' also refers to the geometrical space we take ourselves to experience in the everyday, including the experience of three large dimensions. Of course, quantum field theory, general relativity, and everyday experience describe space in strictly incompatible ways: flat and curved, relativistic and not. But the relations between these descriptions, especially as limits of each other, are well-enough understood to make the notion of a single phenomenal space, described by these phenomenal theories, clear enough for our purposes. 

In contrast, any `space' that appears in the formulation of string theory is `theoretical'; the ultimate question of this paper is of the relation between phenomenal and theoretical spaces, whether they are identical, or whether one is reducible in some sense to the other. Of course, we have seen that the formulation of (quantum) string theory involves two such `theoretical' spaces, \emph{winding space} (explained above), and what I have so far generally called `physical space', in conformity with familiar usage. At this point there are two reasons to adopt a new term -- `\emph{target space}' -- for the latter: first, all the concepts of space we have discussed are `physical' in some general sense; second, on one interpretation of T-duality, target space will turn out to have novel features, and the novel name will avoid the inapt connotations that would come with a pre-existing concept. `Target space' is a term of art in string theory, referring to the `background' space in which a string is embedded: by the function $X(\sigma,\tau)$ at the start of the paper. The classical string is literally located in target space, and wound around it; we will take it that it makes sense to extend this intuitive picture to the quantum string, which thus also `lives' in a background target space. As we have formulated the theory, this situation is represented by the string's position/momentum wavefunction being a field with respect to target space: just as we represent a quantum particle being `in' a region by a wavefunction defined in that region.\footnote{In a more conventional approach to string quantization, $X(\sigma,\tau)$ -- hence location in target space -- is treated as a quantum field.} It is natural when first introduced to string theory, to think that target space is simply the same space we ordinarily experience, or at least space as conceived in contemporary physics. T-duality makes this identification problematic.

Given that general relativity and quantum field theory (and our everyday understanding) are the context of phenomenal space, measurements of its radius are operationalized in their terms: as in Brandenberger and Vafa's thought experiment, for example, which appeals to the photons and clocks of extant physics. Thus the radius of phenomenal space, as I have defined it, is given by $c$ times the duration of the photon's journey. In terms of such measurements, phenomenal space is manifestly very large: we don't know its radius (or even whether it is compact), but we can observe $10^{10}$ light years of it -- and even a simple glance around the room shows that it is much larger that $10^{-33}$m!\footnote{In fact we will count any additional `small', dimensions also as `phenomenal': though they may be required by certain theories of quantum gravity, their possibility is not at all quantum mechanical, as the original Kaluza-Klein theories demonstrate. Even though they are microscopic relative to the ordinary dimensions, they still have a large radius on the relevant scale, $R>\ell_s$.} Moreover, we have also seen how dual theories will both predict that empirical result. While giving dual descriptions of the photon experiment -- one in a target space of the same radius as phenomenal space, and one in a target space (apparently) with the reciprocal radius -- they will agree on its duration, and hence on the radius of phenomenal space. Clearly we cannot immediately infer that target space and phenomenal space are one and the same; the remainder of the paper explores this situation.

At the start of this section we made our first interpretive decision (to be justified below): we decided that dual theories state the same physical facts. Now that we have clearly distinguished three concepts of space -- phenomenal, target and winding -- we are in a position to describe the second interpretive decision, which like the first presents a dichotomy.\\

\noindent \textbf{Interpretation One:} Suppose that the radius of phenomenal space has been measured, by the Brandenberger and Vafa experiment say, and found to be very large. Consider a theory, $T$, that sets $R$, the radius of the $x$-dimension, equal to this observed radius of phenomenal space. One can then understand $T$ in a naively realist way: take $x$ to represent target space, take the string position/momentum wavefunction $\Psi(x,t)$ to represent a string living in target space, and identify target and phenomenal space. That is in fact a very natural way to take the theory. But then how is one to understand the dual theory, $T'$, which we are taking to state the very same physical facts as $T$? Yet, for example, $T$ and $T'$ apparently assign different radii to target space, and (for $w\neq n$) apparently assert that the string is wound a different number of times around target space: these are physically different states of affairs. The solution is to take the duality mapping as specifying a \emph{translation manual}. From (\ref{eq:map}), in the dual theory, let $n$ denote the winding number, not wave number, and $w$ denote the wave number, so that momentum and winding are unchanged! And while in $T$ the $x$-dimension represents target space and the $y$-dimension winding-space, in $T'$ the roles are reversed, so the same wavefunctions pertain to each space as before; and we again identify target and phenomenal spaces so that in $T'$ it is $y$, not $x$, that represents phenomenal space. Finally, as we saw earlier, within each theory the $x$- and $y$-dimensions have reciprocal radii, so in $T'$ the radius of the $y$-dimension is $R/\ell_s^2$: if we understand $T'$ to involve a rescaling of length units by a (dimensionless) factor of $1/\ell_s^2$, then the duals even agree on the radius of target space.\footnote{The factor is dimensionless because the numerator has units of $length^2$. To see that this rescaling is trivial, note that we could have simply have worked in units in which $\ell_s=1$, in which case no rescaling is necessary.} In short, according to this interpretation, duals only appear to be incompatible because they are written in different languages, assigning different meanings to the same words: for example, they appear to assign different radii to target space, but only because they denote different things by `target space'. 

In the framework of elementary formal logic, in this interpretation duals are just related by a permutation of terms that induces a different formal interpretation with respect to a fixed domain, rather than any change in the domain referred to by those terms. Such a permutation is trivially possible for any theory with more than one term (of a given type), so duality requires more. It is an interesting question to say what, and one that I will address carefully in a subsequent paper. In short, continuing to speak in the framework of formal logic, it is not that the set of theorems concerning observables is unchanged: for instance, suppose `$w=1$ and $n=2$' were a theorem in one description, then in the dual `$n=1$ and $w=2$' would be a theorem. But these cannot both be theorems of a single (consistent) system, so the duals have different theorems. Rather, what makes such a permutation a \emph{duality} is that the set of theorems \emph{which only involve observables from some special subset} is invariant. What subset? The ones whose empirical meaning is antecedently fixed. In Brandenberger and Vafa's thought experiment, for instance, both duals agree on the energy of the particle observed, and the duration of the trajectory: in general, as far as the experiment is described in phenomenal terms, the duals agree on the facts. The phenomenal quantities will typically be complicated functions, including classical limits of quantum ones, but theorems concerning them alone will be invariant under T-duality, and that is a non-trivial fact about string theory. \\

\noindent \textbf{Interpretation Two:} The second interpretation of duality also takes the dual theories as asserting all the same things about the physical world, but now under \emph{the same interpretation} of the terms. In this case  what either says about the physical world must be restricted to their shared consequences: for instance, the mass spectrum of the string is common to both and hence a physical fact. Similarly, as we saw, the duals predict that a photon will take the same time to circle the (phenomenal) universe: $10^{12}$ years, say. Since the radius of phenomenal space is thus a shared consequence of the duals, it is a determinate, physical fact.

But the theories do not agree on the radius of target space, nor, as we saw, on what string process corresponds to the photon measurement. Since in this interpretation the terms of the duals denote all the same things, these disagreements are logical incompatibilities between them; and then, because we are taking duals to agree on the physical facts, where the duals disagree, they do not state physical facts. In particular, in this interpretation according to $T$, target space has radius $R$; according to $T'$, the radius is $\ell_s^2/R$. Thus there is no physical fact of the matter which is correct, and with respect to these two values \emph{the radius of target space is indeterminate}. Similarly, it is indeterminate whether the string is wound $w$ or $n$ times around the dimension. And so on.

\citet[\S6]{Matsubara:2013uq} argues along similar lines (as we did in \citealp{Huggett:2013sf}), though he takes (with  misgivings) the shared commitments of the duals to be `structure', which I'm not convinced illuminates these matters without further elaboration.  However, his account does not recognize the role of phenomenal space in the logic of the situation. For instance, at one point (485-6) he correctly says that `space in the mathematical formalism' (i.e., according to $T$ or $T'$ read literally) has unphysical properties, especially determinate radius. And he goes on to infer that `\emph{physical} space' is indeterminate with respect to such properties. Here he is referring to target space as I have understood it, and we are in agreement. But phenomenal space is also physical, and has determinate radius, $R$ in our example. Of course, phenomenal space is derived in some sense, from such processes as those analysed by Brandenberger and Vafa, but that does not mean it is not physical: things reduced to physical things are also physical. Only by ignoring phenomenal space, or incorrectly asserting that only target space is physical, can one reach Matsubara's stated conclusion.

Moreover, because phenomenal space has a determinate radius it follows immediately on this interpretation of duality that phenomenal is space is not identical with target space, whose radius is \emph{in}determinate. Nothing can have a radius that is determinate and indeterminate at once. (Of course the same point applies for theories in which space has multiple dimensions.) Similarly, it follows that we cannot think naively of strings as spatial objects, since there is no fact of the matter (even in a quantum mechanical sense) of how many times they wrap around a dimension. As Brandenberger and Vafa conclude (393), `the invariant notions of general relativity \dots\ may not be invariant notions for string theory'.

If this position seems outr\'e, that is only because it implements perfectly ordinary considerations in a surprising way. Consider Newtonian mechanics: we know that the predictions of the theory are the same whatever point we choose for the origin, whatever orientation we choose for the axes, and indeed whatever constant state of motion we choose for the frame. And so we think that there is no preferred `centre', that space is isotropic, and that spacetime does not distinguish a preferred state of rest. The fact that our coordinates do distinguish a point, break isotropy, and give a notion of rest is quite clearly an artifact of the representation: inertial coordinates make distinctions beyond those we wish to represent. The same understanding can apply to string theory: T-duality shows that a definite radius for target space and a definite state of winding are not physical, but only artifacts of the representation.\footnote{If one follows Witten's \citeyearpar[26]{wit96} suggestion, and interprets $X$ as an internal conformal field, then the picture seems even less remarkable. In private conversation, Matsubara put it to me that analogous indeterminacies arising from other dualities would be harder to swallow: what if the global topology, or dimensionality were indeterminate? I would follow the same line those cases: the upshot is that target `space' has even less spatial structure than we thought, and certainly far less than its representation as a space.}

And of course, once again, there is no mystery about why string theory attributes a determinate radius to phenomenal space, despite the indeterminate radius (on this interpretation) of target space. It is because the radius of phenomenal space is that quantity measured by such experiments as that described by Brandenberger and Vafa. In those cases the low (in string terms) energy of the experiments -- which are the kind of experiments relevant to experience, general relativity and quantum field theory -- guarantees that the result will be determined by the size of the larger of winding or target space, so they agree on the radius of phenomenal space. Duality implies no indeterminacy about that fact.  \\

So we have two interpretational forks. First, do the two theories describe the same physics, or not? And second, if they do, should we take them literally, with the string living in phenomenal space, and avoid incompatibilities by interpreting their terms differently? Or do they have the same formal interpretation, in which case their incompatible assertions imply that there is no physical fact of the matter about the radius of target space, for instance? We will work through the first fork with a simple analogy in the next section; then turn to the second fork in the following section.

\subsection{Interpretation: Physical Equivalence?}

If two theories are dual then under the duality all expectation values are preserved. Thus duality is considerably stronger than `empirical equivalence': it isn't merely that the theories are indistinguishable with respect to a proper subset of their attributions of physical quantities, the `observable' ones. Rather duals are indistinguishable with respect to \emph{all} physical quantities represented by observables. It's not, for instance, that duals have the same predictions for phenomena visible to the naked eye, but might differ on the properties of smaller things -- the agreement is all the way down.

That said, it is true that systems which have dual descriptions in this sense are not necessarily physically equivalent: duality alone is not sufficient for physical equivalence. For example, consider a simple harmonic oscillator, a mass moving horizontally and frictionlessly on a spring, described by the Hamiltonian

\begin{eqnarray}
\label{eqn:SHOH}
H = \frac{p^2}{2m}+\frac{kx^2}{2},
\end{eqnarray}
where $p$ and $x$ are momentum and displacement respectively, and $m$ and $k$ are mass spring constant respectively. This oscillator is dual to another under the duality mapping

\begin{eqnarray}
\label{SHOD}
\nonumber (m,k) & \leftrightarrow & (1/k, 1/m)\\
(x,p) & \leftrightarrow & (p,-x).
\end{eqnarray}
(I will generally say that position is dual to momentum and \textit{vice versa}, although the sign change means that this is not quite accurate. We will pay attention to the sign at the places in which it is significant.) 

That is, as for strings, the Hamiltonian and canonical commutation relations are the same (for the latter, $[x,p]=[p,-x]$). Thus the expectation values for all pairs of dual observables agree as before, and any series of measurements of the quantities represented by quantum observables is consistent with either oscillator, if it is compatible with one. But if the theories describe literal, concrete, physical oscillators in our world, then the  two systems are not the same (assuming that we aren't considering the `self-dual' case, $m=1/k$), and are indeed readily distinguishable, by measuring the masses, for instance: the mass on the first spring is $m$, that on the second $1/k$. There's no conflict in this case between the duality of the theories and their distinguishability, as we can see by exploring the duality more carefully. First we will examine the \emph{in}distinguishability of the duals -- then discuss the distinguishibility.

Quantum theories have classical-number parameters in them, such as the mass here (and the radii of compactified dimensions in string theory), which can be inferred from measurements of the quantities corresponding to the observables of the theory. For instance, in the quantum harmonic oscillator, the energy spectrum is $E_n = \hbar\sqrt{\frac{k}{m}}(n+\frac{1}{2})$, so observations of the energy will determine the c-number $k/m$. But if a pair of theories, with different values for parameters, are dual, then even complete knowledge of expected values for observables, and the formal relations of observables to one another (the commutation algebra, that is) will not determine the correct values.\footnote{Let me emphasise at this point that I am allowing complete epistemic access to the expectation values of every observable -- the value of every quantum quantity -- in every possible state. Clearly that is a lot more than we actually know.} For example, measurements of energy manifestly won't do the job of determining whether $\langle \mathrm{mass},\mathrm{\ spring\ constant}\rangle = \langle m,k \rangle \mathrm{\ or\ } \langle 1/k,1/m \rangle $ -- energy only determines the ratio of mass to spring constant, on which they agree. And the identical patterns of expectation values in the duals means that both possibilities are compatible with any measurements of the observables of the system.

For instance, consider a (partial) analogue of Brandenberger and Vafa's imagined measurement of the radius of the universe. It's worth working through this example briefly just to reinforce those considerations, and to demonstrate their general applicability in cases of duality. First, the energy eigenstates satisfy the time-independent Schr\"odinger equation

\begin{equation}
\label{SHOSE}
\frac{1}{2m}\cdot\frac{\mathrm{d}^2}{dx^2}\psi(x)+\frac{kx^2}{2}\cdot\psi(x) = E\cdot\psi(x),
\end{equation}
(i.e., replace $p$ with the momentum operator in (\ref{eqn:SHOH})). It is a mathematical fact that solutions are either symmetric or antisymmetric with respect to $x\to-x$: $\psi(-x)=\pm\psi(x)$. Hence $|\psi(x)|^2$ is symmetric, $x\cdot|\psi(x)|^2$ is antisymmetric, and $\int\mathrm{d}x\  x\cdot|\psi(x)|^2=0$: i.e., in an energy eigenstate, the expected value of position is 0.

But now consider the dual system, satisfying the dual Schr\"odinger equation (obtained by applying (\ref{SHOD}) to (\ref{SHOSE})):

\begin{equation}
\label{ }
\frac{k}{2}\cdot\frac{\mathrm{d}^2}{dp^2}\phi(p)+\frac{p^2}{2m}\cdot\phi(p) = E\cdot\phi(p).
\end{equation}
The form is unchanged, and so the same considerations apply: the expected value of $p$ is also zero -- of course, this is just a special case of expectation values being preserved by the duality. Now, in the original system $x$ represented physical position, but in the dual system it is $p$: that is determined by the canonical commutation relations. So measurement of the position expectation values does not distinguish the duals: even though each represents the spatial state in a different mathematical space, either as a function of $x$ or of $p$.\footnote{\label{<p>} Making the same points from a different direction, we see an example of how a duality can simplify calculations (something that explains a great deal of the interest in dualities). Suppose I want to know the expected momentum of an oscillator energy eigenstate. The duality tells me that in the dual system the momentum of my oscillator plays the role of position, and I know that the expected position of any oscillator energy eigenstate is zero: so the momentum of mine must be 0.}  And so on for every observable quantity of the theory in every allowed state (every superposition of the energy states, that is): the duals are  indistinguishable by such measurements.

However, for physical oscillators in our world, mass has a physical significance beyond the quantum observables of this formal model. For instance, one could (in principle) dismantle an oscillator and weigh the mass, find out whether it is $m$ or $1/k$. The point is simple: the oscillator is just a subsystem in our world, described by quantities beyond the observables of the model. Hence the models are discernible despite a duality -- by looking outside the model.\footnote{Formally speaking, the identity of an observable is determined by its place in the algebra, and a duality is an isomorphism which preserves expectation values in any state. Looking outside the model amounts to expanding the algebra, for which the duality mapping will not preserve the relations between observables. \citet[479]{Matsubara:2013uq} also points out that real world oscillators are not dual.}

But the case is disanalogous to string theory, if that is taken as a theory of \emph{everything}. What happens if a duality applies to a `total' theory, in the sense that it is the complete physical description of a world, so that there is nothing outside the theory? Specifically, consider a world in which the harmonic oscillator is the complete physical description, so that there is no more encompassing theory by which mass or the spatial amplitude of oscillation could be uniquely determined. What do we now say about the parameters, such as mass, that differentiate the duals? Do they distinguish two distinct physical possibilities, which nevertheless agree on the values of all observables?

To think so would not be  a logical fallacy, nor do I think there are unavoidable semantic or ontological principles that can force the conclusion that the duals describe the same physical possibility. But the case of dual total theories is clearly one in which the putative differences are `hidden' in a very strong sense -- a unique mass is impossible to determine from the physical quantum quantities of the harmonic oscillator, just as a rest frame is from relativistic quantities in special relativity. And when there are quantities that do not supervene on any of the other physical quantities, and when there is no reason to think that different values for them can be determined directly, then at very least from a practical, scientific point of view, it makes sense to treat those differences as non-physical (until some new, well-supported theory shows how they are connected to physical quantities). In other words, long established, well-motivated scientific reasoning should lead us to think that dual total theories represent the same physical situation.

At this point I want to acknowledge that thus assuming that physical states supervene on expectation values is to take a strong stance on the interpretation of QM. For instance, that assumption is clearly false according to Bohm's theory. Moreover, \cite{Nik:07} has shown that Bohmian string theory breaks T-duality as a symmetry. While I take the Bohmian view very seriously indeed, in this discussion we will explore the consequences of duality for interpretations in which there are no `hidden' variables.

Finally, I want to head off the line of thought that we can just see -- immediately experience -- that the radius of space is large, and that things would seem different if it were not. Brandenberger and Vafa's argument applies here. Given that our visual experiences supervene on the physical, whatever physical process that underwrites our experience of a large dimension is realized in both duals: in one as a process involving momentum modes, say, and in the other involving winding modes. I have been arguing that we should take these to be different representations of just one process, but even on the view that counts them as distinct physical possibilities, a fairly mild assumption will guarantee the indistinguishability of the duals even in direct experience. For the two processes will only be experientially distinct if visual experiences depend on the processes grounding them involving spatial (not winding modes): that T-dual brains are not identical minds. It is, in other words to privilege the spatial in the physical theory of mind. But I see no particular motivation for such a view: rejecting it means that dual brains have the same experiences, so that things would not appear any different at all if target space had the reciprocal radius. Hence we cannot just `see' which of the two possibilities holds, and considerations of direct experience provide no reason to think that there are two physical possibilities at all.

\subsection{Interpretation: Factual or Indeterminate Geometry?}

We now proceed on the understanding that T-dual theories describe the same physical situation. The question now is \emph{what} situation that is, in particular with respect to the geometry of space. Above we described two possibilities: it could be that the duals agree that the radius of target space is greater than $\ell_s$, and the apparent inconsistency is resolved by understanding duality as a permutation of terms, a relabeling. Or it could be that the duals should receive the same formal interpretation, so that only their common pronouncements describe what is physical: for instance, a unique radius to phenomenal, but not target, space. In this section I will make a couple of comments about these two possibilities, and then explain why I favor the second.

Talking of `relabeling' the terms of a theory may suggest that the difference is between passive and active interpretations of duality. But that clearly isn't correct: an active transformation links two distinct states of affairs, but both interpretations agree that there is only one possibility, so neither amounts to the view that T-duality is an active transformation. Moreover, T-duality cannot be seen as a passive transformation in the sense that the duals are descriptions of a single situation from two points of view, for the duality does not map `observers' or concrete `reference frames' into distinct but symmetrical observers and frames. And in the looser sense that both interpretations take duals to be distinct representations of the same physical situation, both interpretations take a duality to be equally `passive'.

In fact, the two interpretations that I have described are much closer to the interpretive options that arise in the case of a gauge symmetry. On the one hand, maybe there is `one true gauge' \citep{Hea:01}: in the present context, phenomenal space is identified with target space, and has a definite radius $R>\ell_s$. On the other, maybe apparent differences in choice of gauge are nothing but differences in `surplus representational structure' \citep{Red:75}: target space is distinguished from phenomenal space, and the difference between target spaces of radii $R$ and $\ell_s^2/R$ is merely a difference in representational fluff. We won't pursue this parallel to gauge symmetry in field theory at length, but a couple of points are worth making. First, duality is neither a local nor a continuous symmetry of the kind found in field theory, so much of the philosophical discussion of those theories is inapplicable. Second, that said, at $R=\ell_s$ there is a continuous SU(2)$\times$SU(2) gauge symmetry of which T-duality is a part (e.g., \citealp[247-8]{polchinski2003string}). Thus, in this sense at least, T-duality is formally, and not just conceptually, a gauge symmetry.\footnote{See \citet{Hea:07} and the responses to it for continuous gauge symmetries in general. The SU(2)$\times$SU(2) symmetry entails that an infinitesimal increase of the radius from $R=\ell_s$ is the same as an infinitesimal decrease. \cite{Read:2014fk} makes a related comparison, but between string dualities and diffeomorphism symmetry rather than conventional gauge symmetries.}

So, why do I advocate the indeterminate $R$ interpretation? After all, the definite radius view presented above is intuitive, in that strings live in a space with the phenomenal radius $R>\ell_s$; whether that space is called target or winding space. However, there is a \emph{distinct}, indistinguishable definite radius view  according to which they live in a space whose radius is $\ell_s^2/R$; whether that space is labeled target or winding space. If there is one true gauge, then there are as many distinct possibilities as choices of gauge: in this case two, depending on the radius of the space in which the strings literally live, move and wind. According to one choice, the space of experience is the one in which strings live, while according to the other the space of experience is much bigger than the one in which they live: from Brandenberger  and Vafa we understand how the appearances are the same in either case because of T-duality. The bottom line is that understanding T-duality as a mere permutation of terms leaves open what underlying facts are equally described by the duals -- because it is compatible with different true gauges. It does not really address the issue it was supposed to resolve: dual theories are physically equivalent on this interpretation, but there is a second pair of duals that differs physically from the first, but \emph{only} with respect to an unobservable radius. If one is satisfied with that situation, then why was one not satisfied with physically inequivalent duals?

Moreover, these considerations point to an analogy to related cases in which we usually do accept that there is no fact of some matter (I alluded to a similar example earlier). For instance, one could claim that there is a preferred rest frame in spacetime, even though it has no physical influence in special relativity.  One could even claim that it is some frame which can be picked out physically and phenomenally: for example, perhaps the fixed stars (idealized as an inertial frame) are at rest. I think that the proposal will strike most people as completely unmotivated. But replace `frame' with `radius', and the fixed stars with the phenomenal radius, and the parallel is perfect. Looked at this way, the definite radius view appears as a reactionary attempt to preserve aspects of an old theory when it is superseded, and understood as merely effective.

\section{Conclusions}

The main conclusions of this paper are as follows. First, T-duality is an unusually deep symmetry between theories, with respect to some very counter-intuitive and surprising parameters: especially the radius of space. Gauge symmetries in field theory are similarly deep, but since they typically involve internal degrees of freedom, they are not so shocking. A touchstone of this paper has been the analysis of Brandenberger and Vafa, which explains how there can be two theories  apparently differing on the radius of space, yet predicting the same observed radius. Their analysis has helped at several points to understand the physical meaning of T-duality: such a picture is crucial to understanding duality.

The symmetry is so deep -- between all observables, not just empirical quantities in some superficial sense -- that duals should be understood as giving physically equivalent descriptions. Since they formally disagree on some claims, I have argued (against an alternative view) that the physical commitments of dual theories are limited to their common implications. Specifically, they disagree on the radius of target space, so that must be indeterminate between the two possible values. And in general, `target space' is not a space in the familiar sense at all, but a `space' with only the structures on which the duals agree. (Quite possibly then, a structure that appears as a formal representation of some more fundamental, as yet unknown, non-spatial object.) As the analysis of Brandenberger and Vafa explains, duals do agree on the radius of phenomenal space, so that is determinate. But nothing can be both determinate and indeterminate with respect to radius, and so target space is not phenomenal space. 

Therefore phenomenal space, specifically as a geometric space of determinate radius, is not a fundamental object of string theory, but an appearance, arising from physical processes of the kind that Brandenberger and Vafa analyzed. That, ultimately, is the ontological significance of T-duality.\\

\noindent\textbf{Acknowledgements:} I want to thank Dave Baker, Doreen Fraser, Brian Greene, Jeff Harvey, Keizo Matsubara, Joshua Norton, James Read, Tiziana Vistarini, Christian W\"uthrich, and Eric Zaslow, for help at various stages of this paper; I am also grateful for some feedback from two referees. This work was supported by a Collaborative Research Fellowship from the ACLS.\\

\begin{center}\textbf{References}\end{center}

\bibliographystyle{plainnat}
\bibliography{biblio}

\begin{thebibliography}{19}
\providecommand{\natexlab}[1]{#1}
\providecommand{\url}[1]{\texttt{#1}}
\expandafter\ifx\csname urlstyle\endcsname\relax
  \providecommand{\doi}[1]{doi: #1}\else
  \providecommand{\doi}{doi: \begingroup \urlstyle{rm}\Url}\fi

\bibitem[Bogen and Woodward(1988)]{BogWoo:88}
James Bogen and James Woodward.
\newblock Saving the phenomena.
\newblock \emph{The Philosophical Review}, pages 303--352, 1988.

\bibitem[Brandenberger and Vafa(1989)]{BraVaf:89}
Robert Brandenberger and Cumrun Vafa.
\newblock Superstrings in the early universe.
\newblock \emph{Nuclear Physics B}, 316:\penalty0 391--410, 1989.

\bibitem[Cartwright(1983)]{Car:83}
Nancy Cartwright.
\newblock \emph{How the laws of physics lie}.
\newblock Cambridge Univ Press, 1983.

\bibitem[Dawid(2007)]{Dawid:2007gf}
Richard Dawid.
\newblock Scientific realism in the age of string theory.
\newblock \emph{Physics and Philosophy}, \penalty0 (011), 2007.
\newblock URL \url{http://physphil.uni-dortmund.de}.

\bibitem[Dieks et~al.(forthcoming)Dieks, van Dongen, and de~Haro]{DieDon:15}
Dennis Dieks, Jeroen van Dongen, and Sebastian de~Haro.
\newblock Emergence in holographic scenarios for gravity.
\newblock \emph{Studies in History and Philosophy of Modern Physics},
  forthcoming.

\bibitem[Greene(1999)]{Gre:99}
Brian Greene.
\newblock The elegant universe: Superstrings, hidden dimensions, and the quest
  for the ultimate theory. vintage series, 1999.

\bibitem[Healey(2001)]{Hea:01}
Richard Healey.
\newblock On the reality of gauge potentials.
\newblock \emph{Philosophy of Science}, 68\penalty0 (4):\penalty0 432--455,
  2001.

\bibitem[Healey(2007)]{Hea:07}
Richard Healey.
\newblock \emph{Gauging What's Real}.
\newblock Oxford University Press, 2007.

\bibitem[Huggett and Wuthrich(2013)]{Huggett:2013sf}
Nick Huggett and Christian Wuthrich.
\newblock Emergent spacetime and empirical (in)coherence.
\newblock \emph{Studies in History and Philosophy of Science Part B: Studies in
  History and Philosophy of Modern Physics}, 44\penalty0 (3):\penalty0
  276--285, 2013.

\bibitem[Matsubara(2013)]{Matsubara:2013uq}
Keizo Matsubara.
\newblock Realism, underdetermination and string theory dualities.
\newblock \emph{Synthese}, 190\penalty0 (3):\penalty0 471--489, 2013.

\bibitem[Nikoli{\'c}(2007)]{Nik:07}
Hrvoje Nikoli{\'c}.
\newblock Bohmian mechanics in relativistic quantum mechanics, quantum field
  theory and string theory.
\newblock \emph{Journal of Physics: Conference Series}, 67\penalty0
  (012035):\penalty0 1--6, 2007.

\bibitem[Polchinski(2003)]{polchinski2003string}
Joseph~Gerard Polchinski.
\newblock \emph{String theory}.
\newblock Cambridge university press, 2003.

\bibitem[Read(2014)]{Read:2014fk}
James Read.
\newblock The interpretation of string-theoretic dualities, December 2014.
\newblock URL \url{http://philsci-archive.pitt.edu/11205/}.

\bibitem[Redhead(1975)]{Red:75}
M.~L.~G. Redhead.
\newblock Symmetry in intertheory relations.
\newblock \emph{Synthese}, 32\penalty0 (1-2):\penalty0 77--112, 1975.

\bibitem[Rickles(2011)]{Rickles:2011db}
Dean Rickles.
\newblock A philosopher looks at string dualities.
\newblock \emph{Studies in History and Philosophy of Science Part B},
  42\penalty0 (1):\penalty0 54--67, 2011.

\bibitem[Rickles(2013)]{Rickles:2013ul}
Dean Rickles.
\newblock Mirror symmetry and other miracles in superstring theory.
\newblock \emph{Foundations of Physics}, 43\penalty0 (1):\penalty0 54--80,
  2013.

\bibitem[Teh(2013)]{Teh:2013rc}
Nicholas~J. Teh.
\newblock Holography and emergence.
\newblock \emph{Studies in History and Philosophy of Science Part B: Studies in
  History and Philosophy of Modern Physics}, 44\penalty0 (3):\penalty0
  300--311, 2013.

\bibitem[Witten(1996)]{wit96}
Edward Witten.
\newblock Reflections on the fate of spacetime.
\newblock \emph{Physics Today}, pages 24--30, April 1996.

\bibitem[Zwiebach(2004)]{Zwi:04}
Barton Zwiebach.
\newblock \emph{A first course in string theory}.
\newblock Cambridge university press, 2004.

\end{thebibliography}

\end{document}